\documentclass[prb,twocolumn,showpacs,epsfig,epsf]{revtex4}
\usepackage{amsmath,amsthm,amsfonts,amscd,mathrsfs,pifont,bm}
\usepackage{eucal,graphicx}
\begin{document}

\title  {Integrability of zero-dimensional replica field theories at $\beta=1$}

\author {Pedro Vidal${}^{1,2}$ and Eugene Kanzieper${}^{1}$}

\affiliation
       {
       ${}^1$ Department of Applied Mathematics, H.I.T.---Holon Institute of
       Technology, Holon 5810201, Israel \\
       ${}^2$ Fakult\"at f\"ur Physik, Universit\"at Bielefeld, Bielefeld 33615, Germany
       }
\date   {September 5, 2013}

\begin  {abstract}
Building on insights from the theory of integrable lattices, the integrability is claimed for nonlinear replica $\sigma$ models derived in the context of real symmetric random matrices. Specifically, the fermionic and the bosonic replica partition functions are proven to form a single (supersymmetric) Pfaff-KP hierarchy whose replica limit is shown to reproduce the celebrated nonperturbative formula for the density-density eigenvalue correlation function in the infinite-dimensional Gaussian Orthogonal Ensemble. Implications of the formalism outlined are briefly discussed.
\end{abstract}

\pacs   {05.40.-a, 02.50.-r, 11.15.Ha, 75.10.Nr \qquad\qquad\qquad\qquad Phys. Rev. E {\bf 88}, R-030101 (2013)}
\maketitle
\newpage
{\it Introduction.}---Perhaps the most comprehensive and unifying quantitative description of disordered and quantum chaotic systems can be achieved by using field-theoretic methods. Since the seventies of the last century, a number of such field-theoretic frameworks have been devised. Having different mathematical status and physical areas of applicability, they all rest on the concept of functional field integrals and are known as nonlinear $\sigma$ models \cite{W-1979,ELK-1980,F-1983,E-1982a,VWZ-1985,HS-1990}.

In the context of disordered systems, three major formulations of
nonlinear $\sigma$ models exist: the replica
\cite{W-1979,ELK-1980,F-1983}, the supersymmetry
\cite{E-1982a,VWZ-1985}, and the Keldysh
\cite{HS-1990} $\sigma$ models. Of these, replicas is the earliest \cite{EA-1975,W-1979,ELK-1980} and the most controversial invention that continues to challenge \cite{VZ-1985,Z-1999,T-2003} both mathematical and theoretical physicists. Decades after debuting in condensed matter physics, an operationally transparent and controllable treatment of replica field theories is barely available \cite{K-2011} even in the mathematically simplest yet physics motivated setup provided by the random matrix theory \cite{M-2004} (RMT). Indeed, it took nearly twenty years to extract some nonperturbative results out of replicas,
in both the RMT limit
\cite{KM-1999a} and beyond it \cite{KM-1999b}. More precisely, adopting the ideas of replica symmetry breaking originally devised in the theory of spin glasses \cite{T-2003}, the authors of Ref. \cite{KM-1999a} managed to show that the {\it fermionic} variation \cite{ELK-1980} of replicas is capable of producing partially nonperturbative RMT results for the microscopic two-point density-density spectral correlation function
\begin{eqnarray} \label{ddcf-asymp}
    R_2^{(\beta)}(\omega) \simeq -\frac{1}{\beta (\pi \omega)^2} &+& \frac{2 \Gamma^2(1+2/\beta)}{(2\pi\omega)^{4/\beta}}\, \cos(2\pi\omega) \nonumber\\
    &+& \frac{\delta_{\beta,4}}{2 (2\pi \omega)^4} \cos(4\pi\omega).
\end{eqnarray}
Equation (\ref{ddcf-asymp}) describes two-point eigenvalue correlations in `infinite-dimensional' real symmetric ($\beta=1$), complex Hermitean ($\beta=2$), or quaternion-real self-dual ($\beta=4$) random matrices whose spectrum was unfolded so as to make the mean level spacing unity, $\Delta=1$. On the formal level, $R_2^{(\beta)}(\omega)$ corresponds to the thermodynamic limit
\begin{eqnarray} \label{rescaling}
    R_2^{(\beta)}(\omega) = \lim_{N \rightarrow \infty} \frac{R_2^{(\beta)}(E_1, E_2;N)}{R_1^{(\beta)}(E_1;N) R_1^{(\beta)}(E_2;N)}
\end{eqnarray}
where $E_{1,2}=E \pm \omega R_1^{(\beta)}(E;N) /2$, and a set of finite--$N$ eigenvalue correlation functions is defined as follows:
\begin{eqnarray}
    R_p^{(\beta)}({\bm E};N) = \left<
    \prod_{\alpha=1}^p
        {\rm Tr\,}\delta(E_\alpha - {\bm {\mathcal H}})\right>_{{\rm G}\beta{\rm E}_N}.
\end{eqnarray}
Here, ${\bm E}$ stands for a set of energy variables, ${\bm E}=(E_1,\cdots,E_p)$, whilst angular brackets $\left<\cdots\right>$ denote averaging over the Gaussian orthogonal ($\beta=1$), unitary ($\beta=2$), or symplectic ($\beta=4$) ensemble of random matrices specified by the normal probability density $\propto \exp(-{\rm Tr\,}{\bm {\mathcal H}}^2)$ on the space ${\mathcal M}_N(\mathbb{F}_\beta)$ of Hermitean $N\times N$ matrices with real $({\mathbb F}_1 = {\mathbb R})$, complex $({\mathbb F}_2 = {\mathbb C})$, or quaternion-real $({\mathbb F}_4 = {\mathbb H})$ entries.

The validity of Eq.~(\ref{ddcf-asymp}), summarising achievements of the early {\it fermionic} replica research, is restricted \cite{Remark-DH} to the energy scales in excess of the mean level spacing, $\omega \gg \Delta = 1$. Notably, several attempts \cite{VZ-1985,Z-1999} to reproduce the same result from {\it bosonic} replicas did not bear fruit. Yet, within the supersymmetry technique \cite{E-1982a}, the same task of computing the density-density correlation function in the RMT setting was accomplished \cite{E-1982a,E-1982b} in less than a year and with greater rigour. For the Dyson $\beta=1$ symmetry class (which will be a focus of this Rapid Communication), the truly nonperturbative eigenvalue density-density correlation function reads \cite{E-1982b,D-1962} ($\omega >0$):
\begin{eqnarray}\label{er}
    R_2^{(1)} (\omega) &=& \delta(\omega) +1 -
    \left(\frac{\sin(\pi\omega)}{\pi \omega}\right)^2 \nonumber\\
    &-& \frac{\partial}{\partial\omega} \left(
    \frac{\sin(\pi\omega)}{\pi \omega}
    \right) \int_\omega^\infty dt\, \frac{\sin(\pi t)}{\pi t}.
\end{eqnarray}
Sadly, heuristic approaches to nonlinear replica $\sigma$ models have so far failed to produce the result of such generality.

One of the reasons behind this misfortune of replicas is
that they are much more quality demanding in making various
approximations and, consequently, are more involved
operationally-wise. Such a vulnerability can be attributed to the
{\it continuous} geometry underlying replica field theories. The latter derives from the very idea of constructing replica generating functionals
which associates a physical observable of interest with the $n\rightarrow 0$ limit of a collective matrix field ${\bm {\mathcal Q}}$ arising as a result of imaginary $n$--fold replication of original physical system. Since the dimension of ${\bm {\mathcal Q}}$ is an integer, ${\rm dim}({\bm {\mathcal Q}})\propto n$, taking the replica limit becomes quite an adventure \cite{VZ-1985,KM-1999a,Z-1999,K-2011}. For instance, the $p$--point Green function
\begin{eqnarray}
    G_p^{(\beta)}({\bm E};N) =
    \left<
    \prod_{\alpha=1}^p
        {\rm Tr\,}\left(E_\alpha - {\bm {\mathcal H}}\right)^{-1}\right>_{{\rm G}\beta{\rm E}_N}
\end{eqnarray}
can formally be recovered from the replica limit
\begin{eqnarray}\label{rL}
    G_p^{(\beta)}({\bm E};N) = \lim_{n\rightarrow \pm 0}
    \frac{1}{n^p} \, \frac{\partial^p}{\partial_{E_1} \cdots \partial_{E_p}}
    \tilde{Z}_{n}^{(\beta)}({\bm E};N),
\end{eqnarray}
where $\tilde{Z}_{n}^{(\beta)}({\bm E};N)$ is either fermionic ($n>0$) or bosonic ($n<0$) representation of the replica partition function
\begin{equation} \label{rpf-def}
    Z_{n}^{(\beta)}({\bm E};N) = \left<
    \prod_{\alpha=1}^p {\rm det}^n(E_\alpha - {\bm {\mathcal H}})
    \right>_{{\rm G\beta E}_N}.
\end{equation}
The two representations can be derived from Eq.~(\ref{rpf-def}) by using standard field-theoretic methods \cite{ELK-1980,VZ-1985,KM-1999a}. Alternatively, one may appeal to the RMT duality relations \cite{D-2009} which yield, e.g., the fermionic replica partition function ($n>0$) in the form
\begin{eqnarray}\label{frp}
    \tilde{Z}_{|n|}^{(\beta,+)}({\bm E};N) &\propto&
     e^{(2/\beta){\rm Tr} [{\bm E}\otimes \openone_n]^2}
    \int [D{\bm {\mathcal Q}}]\, e^{
        -(\beta/2)\,{\rm Tr\,}{\bm {\mathcal Q}}^2}
         \nonumber\\
        &\times&  {\rm det}^N {\bm {\mathcal Q}}\,\, e^{
         - 2i {\rm Tr} \left[ {\bm {\mathcal Q}} ({\bm E}\otimes \openone_n) \right]}.
\end{eqnarray}
Here, an integral runs over the matrix field ${\bm {\mathcal Q}} \in {\mathcal M}_{np}({\mathbb F}_{4/\beta})$ of dimension ${\rm dim} ({\bm {\mathcal Q}})=np$. The bosonic replica partition function ($n<0$) whose derivation is more intricate and delicate \cite{Z-2006} also admits a dual matrix representation \cite{D-2009,Z-2006,F-2002,FK-2003} akin to Eq.~(\ref{frp}).

The necessity of performing a limiting procedure in the matrix dimension [Eqs.~(\ref{rL}) and (\ref{frp})] highlights an unusual continuous geometry of replica field theories; it ruins a classic notion of matrix, raising both conceptual
problems \cite{P-2003} regarding mathematical foundations of nonlinear replica $\sigma$ models and operational problems of
dealing with weird objects where the plain intuition often refuses to work.

To avoid running into an underdeveloped concept of spaces of rational dimension \cite{P-2003}, one first performs replica calculations for an integer number of replicas in a hope to implement the replica limit $n\rightarrow 0$ at a later stage after seeking an analytic continuation away from $n$ integers. For such a continuation to be grounded, one should start with an {\it exact} integer--$n$ result. The latter had been unavailable in the early replica studies \cite{KM-1999a} that heavily relied on existence of a large spectral parameter $\omega/\Delta \gg 1$ required to justify {\it approximate} saddle-point calculation of the replica partition function. The above inequality restricted domain of applicability of Eq.~(\ref{ddcf-asymp}), while an approximate character of integer--$n$ calculations led to its questioned \cite{Z-1999} mathematical status.

The status of nonlinear replica $\sigma$ models was dramatically uplifted in Refs. \cite{K-2002,K-2005,OK-2007} where both {\it fermionic} and {\it bosonic} versions of zero-dimensional replica field theories were shown to be exactly solvable for $\beta=2$ symmetry class. The theory of Painlev\'e transcendents \cite{N-2004}, the Toda lattice \cite{DKJM-1983} and associated $\tau$ functions \cite{AvM-2001} lied at the heart of the exact approach to replicas in the elaboration \cite{K-2002,K-2005,OK-2007}. Later, a complementary -- {\it supersymmetric} -- formulation of replicas was introduced by Splittorff and Verbaarschot \cite{SV-2003-2004}. These authors have argued that the replica limit can efficiently be implemented on the level of supersymmetric Toda lattice equation whose positive and negative branches describe fermionic and bosonic partition functions, respectively. Supersymmetric replicas have greatly simplified calculations of $\beta=2$ spectral correlation functions through a remarkable fermionic--bosonic factorisation \cite{SV-2003-2004,OK-2010}.

In this Rapid Communication, we take one more step towards comprehensive understanding of integrable structures of nonlinear replica $\sigma$ models by presenting an integrable theory of a supersymmetric variation of zero-dimensional replica field theories for $\beta=1$ symmetry class that so far denied a nonperturbative treatment. Focussing on the Gaussian orthogonal ensemble (GOE) and building on the theory of Pfaff--KP $\tau$ functions \cite{AvM-2001}, we shall show that supersymmetric replicas do produce a nonperturbative result Eq.~(\ref{er}) for the GOE two-point density-density correlation function.

{\it Density--density correlation function from supersymmetric replicas.}---Specifying the ${\rm GOE}_N$ by the normal probability density~$\propto\exp(-N {\rm Tr\,} {\bm {\mathcal H}}^2)$ on ${\mathcal M}_N({\mathbb F}_1)$, we look for a large--$N$ limit of the replica partition function $Z_n^{(1)}(E_1,E_2; N)$ whose energy variables $E_{1,2}$ are rescaled as described beneath Eq.~(\ref{rescaling}). A sometwhat lengthy calculation \cite{VK-2013} based on the methods described in Ref. \cite{FK-2003,Z-1996,AvM-2004} shows that both fermionic and bosonic partition functions admit the large--$N$ factorization
\begin{eqnarray}
    {\tilde Z}_{|n|}^{(\pm)}(\omega;N) =
     N^{2n^2} (4e)^{\mp |n|N} \cdot {\hat z}_{|n|}^{(\pm)}(\omega),
\end{eqnarray}
where ${\hat z}_{m}^{(+)} (\omega)$ is the fermionic partition function,
\begin{equation} \label{fpp}
{\hat z}_{m}^{(+)} (\omega) = c_{m}^{(+)} \prod_{k=1}^{m} \int_{-1}^{1} d\lambda_k\, (1- \lambda_k^2)
    \, e^{-i \pi\omega \lambda_k} \,|\Delta_{m}^{\bm \lambda}|^4
\end{equation}
whilst ${\hat z}_{m}^{(-)}(\omega)$ is the bosonic one,
\begin{equation} \label{bpp}
{\hat z}_{m}^{(-)} (\omega) = c_{m}^{(-)} \prod_{k=1}^{2m} \int_1^\infty \frac{d\lambda_k}{\sqrt{\lambda_k^2-1}}
    \, e^{i (\pi\omega/2)\lambda_k}  \,
    |\Delta_{2m}^{\bm \lambda}|.
\end{equation}
Here, $\omega \mapsto \omega+i0$ is assumed to have an infinitesimally small positive imaginary part, $\Delta_m^{\bm \lambda} =\prod_{j>k}^m (\lambda_j-\lambda_k)$ is the Vandermonde determinant,
\begin{equation}
    c^{(+)}_{m} = \frac{(2\pi)^m G(1/2)}{2^{4m^2} m!}
    \frac{\Gamma(m+1/2)}{G(2m+3/2)}
    \prod_{j=1}^m \frac{\Gamma(2m+2j)}{ \Gamma^3(2j)}
\end{equation}
and
\begin{eqnarray}
    c^{(-)}_m = \frac{\pi^m}{2^{2m^2} (2m)!}
    \frac{1}{\prod_{j=1}^{2m} \Gamma^2(j/2)}.
\end{eqnarray}
The notation $G(z)$ stands for the Barnes $G$-function.

In order to recover the two-point density-density correlation function [Eq.~(\ref{rescaling})] in infinite-dimensional GOE,
$R_2(\omega) =(1/2\pi^2)\, {\mathfrak R}e \, g(\omega) - 1/2$, through the replica limit for the corresponding two-point Green function
\begin{eqnarray}\label{g-replica}
    g(\omega) = - \lim_{n\rightarrow 0} \frac{1}{n^2} \frac{\partial^2}{\partial \omega^2} \, {\hat z}_{|n|}^{(\pm)} (\omega),
\end{eqnarray}
one has to find a proper nonperturbative representation of either fermionic \cite{K-2002} or bosonic \cite{OK-2007} partition functions, or opt for supersymmetric replicas \cite{SV-2003-2004}. It is the latter route that will be explored below.

To make a forthcoming presentation transparent, we first quote our main analytic result. Let $\hat{z}_n (\omega)$ be a supersymmetric replica partition function,
\begin{eqnarray}\label{susy-rpf}
    \hat{z}_n (\omega) =
    \left\{
      \begin{array}{ll}
        \hat{z}_{|n|}^{(-)}(\omega), & \hbox{$n<0$;} \\
        1, & \hbox{$n=0$;} \\
        \hat{z}_{|n|}^{(+)}(\omega), & \hbox{$n>0$.}
      \end{array}
    \right.
\end{eqnarray}
For all $n$ integers (both positive and negative), the following nonlinear recursive differential equation of the Pfaff-KP type holds:
\begin{widetext}
\vspace{-0.5cm}
\begin{eqnarray} \label{PfKP-difference}
    \left( \frac{\partial^3}{\partial \omega^3} - \frac{2n}{\omega} \frac{\partial^2}{\partial \omega^2} + \frac{2n}{\omega^2} \frac{\partial}{\partial \omega} \right)\log \hat{z}_n (\omega) &+&
     2 \left(
         \frac{\partial}{\partial \omega} \log \hat{z}_n (\omega)
    \right) \left(
         \frac{\partial^2}{\partial \omega^2} \log \hat{z}_n (\omega)
    \right) \nonumber\\
    &=&
    \pi^4 n^2 (2n+1)\,\omega\, \frac{\hat{z}_{n-1} (\omega)\,\hat{z}_{n+1} (\omega)}{\hat{z}_n^2 (\omega)}
    \left(
     4n + \omega  \frac{\partial}{\partial \omega}
      \log \frac{\hat{z}_{n+1} (\omega)}{\hat{z}_{n-1} (\omega)}
    \right).
\end{eqnarray}
\end{widetext}
Equation (\ref{PfKP-difference}), whose derivation will be sketched below, is central to implementing the replica limit Eq.~(\ref{g-replica}). Indeed, assuming that Eq.~(\ref{PfKP-difference}) stays valid for $n$ real, we introduce a small--$n$ expansion
\begin{eqnarray}\label{z-exp}
    \log \hat{z}_n(\omega) = n f_1(\omega) + n^2 f_2(\omega) + {\mathcal O}(n^3)
\end{eqnarray}
and substitute it into Eq.~(\ref{PfKP-difference}) to realize that $f_1^{\prime\prime}(\omega)$ should be set to zero \cite{Remark-F1} to ensure existence of the limit Eq.~(\ref{g-replica}). Further, we make use of the boundary conditions for $\hat{z}_{|n|}^{(-)}(\omega)$ at $\omega\rightarrow \infty$ [see Eq.~(\ref{bpp})] to figure out that $f_1^\prime(0)=-i\pi$. Finally, we combine Eqs.~(\ref{g-replica}) and (\ref{z-exp}) to observe the relation $g^\prime(\omega) = - f_2^{\prime\prime\prime}(\omega)$ and subsequently derive from Eq.~(\ref{PfKP-difference}):
\begin{equation} \label{g-prime}
    g^{\prime}(\omega) = -\frac{2i\pi}{\omega^2} + \pi^4 \omega^2
    \hat{z}_1^{(+)} (\omega)\, \hat{z}_1^{(-)} (\omega) \frac{\partial}{\partial \omega} \log \frac{\hat{z}_1^{(-)} (\omega)}{\hat{z}_1^{(+)} (\omega)}.
\end{equation}
Hence, the Green function can solely be expressed in terms of replica partition function for one fermionic [$\hat{z}_1^{(+)} (\omega)$] and one bosonic [$\hat{z}_1^{(-)} (\omega)$] flavor. Such a `factorization property' was first observed in Ref. \cite{SV-2003-2004} in the context of supersymmetric replicas for $\beta=2$ matrix models.

To complete the evaluation of $R_2(\omega)$, we use Eqs.~(\ref{fpp}) and (\ref{bpp}) to find out
\begin{eqnarray}\label{z1p}
    \hat{z}_1^{(+)} (\omega) &=& - \frac{4}{\pi^2\omega} S^\prime(\omega), \\
    \label{z1m}
    \hat{z}_1^{(-)} (\omega) &=& \frac{i}{2(\pi \omega)} \int_1^\infty \frac{dt}{t} \, e^{i (\pi \omega) t},
\end{eqnarray}
where $S(\omega) = \sin(\pi\omega)/(\pi\omega)$. Equations (\ref{g-prime}) -- (\ref{z1m}), imply
\begin{eqnarray}
    R_2^\prime (\omega) &=& \frac{1}{\pi} \frac{\partial}{\partial \omega} {\mathfrak R}e\, \frac{i}{\omega+i0} \nonumber\\
    &-& \frac{\partial}{\partial \omega} \left(
        S^2(\omega) + S^\prime(\omega) \int_\omega^\infty dt\, S(t)
    \right)
\end{eqnarray}
or, equivalently,
\begin{equation}
    R_2 (\omega) - R_2(\infty) = \delta(\omega)
    -  \left(
        S^2(\omega) + S^\prime(\omega) \int_\omega^\infty dt\, S(t)
    \right).
\end{equation}
Setting $R_2(\infty)$ to unity concludes our replica derivation of the celebrated nonperturbative formula [Eq.~(\ref{er})] for the two-point density-density correlation function in infinite--dimensional GOE.

{\it Supersymmetric Pfaff--KP equation.}---Since the derivation of the supersymmetric Pfaff--KP equation [Eq.~(\ref{PfKP-difference})] is quite tedious, below we only sketch its main idea leaving the details for a separate publication. 
We proceed in four steps.

{\it (i)}---First, we define the fermionic and the bosonic $\tau$ functions given by
\begin{equation}\label{ftf}
    {\hat \tau_{2m}}^{(+)} (s; {\bm t}) = \frac{1}{m!}
    \prod_{k=1}^{m} \int_{-1}^{1} d\lambda_k\, (1- \lambda_k^2)
    \, e^{2 s \lambda_k + 2 V({\bm t};\lambda_k)}
    |\Delta_{m}^{\bm \lambda}|^4,
\end{equation}
and (${\mathfrak R}e\, s >0$)
\begin{equation} \label{btf}
    {\hat \tau_{2m}}^{(-)} (s; {\bm t}) = \frac{1}{(2m)!}
    \prod_{k=1}^{2m} \int_{1}^{\infty} \frac{d\lambda_k}{\sqrt{\lambda_k^2-1}}
    \, e^{- s \lambda_k - V({\bm t};\lambda_k)}
    |\Delta_{2m}^{\bm \lambda}|,
\end{equation}
respectively. Here, $s = -i\pi\omega/2$, whilst $V({\bm t};\lambda) = \sum_{j=1}^\infty t_j \lambda^j$ is a deformation potential parameterized by infinitely many parameters ${\bm t} =(t_1, t_2,\dots)$. To make connection with the supersymmetric replica partition function ${\hat z}_n(\omega)$, we also construct the supersymmetric $\tau$ function
\begin{eqnarray} \label{susy-tf}
    {\hat \tau_{2m}} (s;{\bm t}) =
    \left\{
      \begin{array}{ll}
        {\hat \tau_{2|m|}}^{(-)}(s;{\bm t}), & \hbox{$m<0$;} \\
        1, & \hbox{$m=0$;} \\
        {\hat \tau_{2|m|}}^{(+)}(s;{\bm t}), & \hbox{$m>0$}
      \end{array}
    \right.
\end{eqnarray}
such that the following projection relation holds:
\begin{eqnarray}\label{proj-relation}
    \hat{z}_n (\omega) = {\hat \tau}_{2n} (s = -i\pi\omega/2;{\bm 0}).
\end{eqnarray}

{\it (ii)}---Second, we utilize the formalism by Adler and van Moerbeke \cite{AvM-2001} to claim the existence of a single `supersymmetric' Pfaff--KP hierarchy coherently constraining both fermionic [Eq.~(\ref{ftf})] and bosonic [Eq.~(\ref{btf})] $\tau$ functions. Its first [${\rm Pf\,KP}_1$] and second [${\rm Pf\,KP}_2$] equations, respectively, read \cite{VK-2013}:
\begin{widetext}
\vspace{-0.5cm}
\begin{eqnarray} \label{pkp-01}
    \left(
        \frac{\partial^4}{\partial t_1^4} + 3\,\frac{\partial^2}{\partial
        t_2^2} -
        4\, \frac{\partial^2}{\partial t_1 \partial t_3}
    \right)\, \log {\hat \tau}_{2m}
     + 6\, \left(
        \frac{\partial^2}{\partial t_1^2}\, \log {\hat \tau}_{2m}
    \right)^2  &=& 12\frac{{\hat \tau}_{2m-2}\, {\hat \tau}_{2m+2}}{{\hat \tau}_{2m}^2},\qquad \\
\label{pkp-02}
    \left(
        \frac{\partial^4}{\partial t_1^3 \partial t_2} - 3 \frac{\partial^2}{\partial t_1 \partial t_4}
        + 2 \frac{\partial^2}{\partial t_2 \partial t_3}
    \right) \,\log {\hat \tau}_{2m} +  6 \, \left(
    \frac{\partial^2}{\partial t_1^2} \log {\hat \tau}_{2m}
    \right) \left(
    \frac{\partial^2}{\partial t_1 \partial t_2} \log {\hat \tau}_{2m}
    \right)  &=& 6 \frac{{\hat \tau}_{2m-2}\, {\hat \tau}_{2m+2}}{{\hat \tau}_{2m}^2}  \frac{\partial}{\partial t_1} \log \frac{{\hat \tau}_{2m+2}}{{\hat \tau}_{2m-2}}.\qquad
\end{eqnarray}
\end{widetext}
Nonlinear differential operators in the l.h.s.~of~Eqs.~(\ref{pkp-01}) and (\ref{pkp-02}) are known as the first and the second Kadomtsev--Petviashvili operators \cite{OK-2010}. Equations (\ref{pkp-01}) and (\ref{pkp-02}) are the $\beta=1$ analogues of supersymmetric Toda Lattice equations previously discovered \cite{SV-2003-2004,OK-2010} in the context of $\beta=2$ replica field theories.

{\it (iii)}---Third, given the invariance of ${\hat \tau}_{2m}^{(\pm)}$ under the change of integration variables $\lambda_k \mapsto \mu_k + \epsilon (\mu_k^2-1)\mu_k^{q+1}$ in Eqs.~(\ref{ftf}) and (\ref{btf}), we observe \cite{VK-2013} that the supersymmetric $\tau$ function Eq.~(\ref{susy-tf}) satisfies an infinite set of Virasoro constraints ($q \ge -1$)
\begin{eqnarray}\label{vcs}
    \Big[ \hat{\mathcal L}_{q+2}^{(1)}({\bm t}) - \hat{\mathcal L}_{q}^{(1)}({\bm t}) + s \left(
        \frac{\partial}{\partial t_{q+3}}
         - \frac{\partial}{\partial t_{q+1}}\right)\qquad\nonumber\\
          - (q+2) \frac{\partial}{\partial t_{q+2}}\Big]\, {\hat \tau}_{2m}(s; {\bm t}) =0,
\end{eqnarray}
where
\begin{equation}
    \hat{\mathcal L}_q^{(1)} = \sum_{j=1}^\infty jt_j \frac{\partial}{\partial t_{q+j}} + \frac{1}{2} \sum_{j=0}^q \frac{\partial^2}{\partial t_j \partial t_{q-j}} + \frac{1}{2} (q+1) \frac{\partial}{\partial t_q}
\end{equation}
is the $\beta=1$ Virasoro operator \cite{AvM-2001}. Equation (\ref{vcs}) assumes that $\partial/\partial t_1 = \partial/\partial s$; the operator $\partial/\partial t_0$ should be interpreted as $\partial/\partial t_0=2m$.

{\it (iv)}---Fourth, we combine Eq.~(\ref{proj-relation}) with the three lowest Virasoro constraints ($q=-1$, $0$, and $+1$) to project $\left[({\rm Pf\,KP}_1) + (s/2n)({\rm Pf\,KP}_2)\right]\,\log \,{\hat \tau}_{2n}(s;{\bm t})$ onto the hyperplane ${\bm t}={\bm 0}$. Lengthy but straightforward calculations yield \cite{VK-2013} the sought Pfaff--KP equation for the supersymmetric replica partition function $\hat{z}_{n}(\omega)$. This completes our derivation of Eq.~(\ref{PfKP-difference}).

{\it Conclusions.}---In this Rapid Communication, we have shown how the ideas of integrability can be utilized to formulate a nonperturbative theory of $\beta=1$ zero-dimensional replicas in their supersymmetric elaboration. Although a particular emphasis was placed onto the GOE random matrices, the formalism outlined is quite general and should equally apply to other random matrix models (including those appearing in various RMT formulations of quantum chromodynamics), for both $\beta=1$ and $\beta=4$ symmetry classes which are dual to each other. Yet, we believe that an understanding of formal structures lurking behind zero-dimensional replica $\sigma$ models that was accumulated during the past decade takes us one step closer to formulating exact replica theories for more realistic matrix models (e.g., random banded matrices) and physical systems. In connection to the latter, we wish to mention a recent breakthrough \cite{CDR-D-2010,CD-2011-D-2012} in rigorization \cite{BC-2011-2013} of heuristic replicas devised in the context of one-dimensional disordered polymers. Concerned with the statistics of free energy fluctuations of a directed polymer, the authors of Refs. \cite{CDR-D-2010,CD-2011-D-2012,GD-2012} used a Bethe Ansatz solution of a replicated system of attractive bosons \cite{K-1987} to derive a set of Tracy-Widom laws for the free energy distribution. Interestingly, while the formalism developed in Refs. \cite{CDR-D-2010,CD-2011-D-2012} is operationally different from ours [see also Refs. \cite{K-2002,SV-2003-2004,OK-2007}], the underlying concept of both approaches -- integrability of a corresponding replicated system -- appears to be precisely the same. Given this observation, will it be possible to establish a formal correspondence between the two replica frameworks?

This work was supported by the Israel Science Foundation through the grants No 414/08 and No 647/12.

\end{document}